\def\square{\,\hbox{\vrule\vbox{\hrule\phantom{N}\hrule}\vrule}\,}
\parindent=0pt 
 
\centerline{\bf On the  structure of the new electromagnetic conservation laws.}

\

\

\centerline{\bf S. Brian
Edgar, }

\centerline{Department of Mathematics, }

\centerline{Link\"{o}pings universitet,}

\centerline{Link\"{o}ping,}

\centerline{Sweden S-581 83.}
\smallskip
\centerline{ email: bredg@mai.liu.se}

\

\

{\bf Abstract.} 
New electromagnetic conservation laws have recently been proposed: in the absence of
electromagnetic currents, the trace of the Chevreton superenergy tensor, $H_{ab}$  is
divergence-free in four-dimensional (a) Einstein spacetimes for test fields, (b) 
Einstein-Maxwell spacetimes. Subsequently it has been pointed out, in analogy with flat
spaces, that  for Einstein spacetimes  the trace of the Chevreton superenergy tensor
$H_{ab}$ can be rearranged in the form of a generalised wave operator $\square_L$ acting on
the energy momentum tensor $T_{ab}$ of the test fields, i.e., $H_{ab}=\square_LT_{ab}/2$.  In
this letter we show,  for Einstein-Maxwell spacetimes in the full
non-linear theory, that, although,  the trace of the Chevreton superenergy tensor
$H_{ab}$ can again be rearranged in the form of a generalised wave operator $\square_G$ acting
on the electromagnetic energy momentum tensor, in this
case the result is also crucially dependent on Einstein's equations; hence we argue that the
divergence-free property of the  tensor 
$H_{ab}=\square_GT_{ab}/2$  has significant independent content
beyond that of the divergence-free property of
$T_{ab}$.

\

\


\

\

It is easy to see that the energy-momentum tensor of source-free electromagnetic fields
(notation and conventions from [1])
$$T_{ab}= F_{ac}
F_b{}^c - g_{ab}F_{cd}
F^{cd}/4
\eqno(1)$$
is divergence-free as a consequence of the source-free Maxwell equations in {\it arbitrary
$n$-dimensional spaces.}

 Bergqvist, Eriksson and
Senovilla (BES) [2] have recently found that the trace of the Chevreton tensor [3],[4], 
 $$H_{ab}\equiv\nabla^{e}F_{ac}\nabla_{e}
F_b{}^c - g_{ab}\nabla^{e}F_{cd}\nabla_{e}
F^{cd}/4 \eqno(2)$$  is 
divergence-free  in {\it four} dimensions for:  

(a)  {\it Einstein  spaces}, with the
source-free electromagnetic field  considered as a test field so that the geometry of the
space is unnaffected; 

(b)  {\it  source-free Einstein-Maxwell spaces} (including  a possible
cosmological constant). 

Subsequently, Deser [5]  has  rederived (the Ricci-flat space version of) part (a) of the BES
result by showing, in {\it four-dimensional Ricci-flat spaces}  that 
$H_{ab}$ can be rearranged into the form $$2H_{ab}=\square_L T_{ab}\equiv
\nabla_i \nabla^i T_{ab} -2 C_{aibj}T^{ij}\ ,\eqno(3)$$ and
then     exploiting the commutativity property  
$$\bigl[\square_L,\nabla^b \bigr]X_{ab}=0\eqno(4)$$ which is valid for {\it arbitrary}
symmetric $2$-index tensors $X_{ab}$, where
$\square_L$
 is the Lichnerowicz generalised wave operator [6] in Ricci-flat
spaces\footnote{${}^{\dag} $}{Lichnerowicz [6] has defined a generalised wave operator
acting on an arbitrary   tensor in {arbitrary spaces}.  In particular, for {\it arbitrary}
symmetric $2$-index tensors in {\it arbitrary spaces} it has the form  
$\Delta X_{ab} \equiv -\nabla_i\nabla^i X_{ab} - R_{ai}X^i{}_b - R_{bi}X^i{}_a
+2R_{aibj}X^{ij}$; Lichnerowicz has shown that in spaces satisfying $\nabla_cR_{ab}=0$ there
exists the identity 
$$\bigl[\Delta,\nabla^b \bigr]X_{ab}=0$$
where $\Delta Y_{a} \equiv -\nabla_i\nabla^i Y_{a} - R_{ai}Y^i $.
We shall reverse the sign and define
$\square_L X_{ab} \equiv
\nabla_i\nabla^i X_{ab} + R_{ai}X^i{}_b + R_{bi}X^i{}_a -2R_{aibj}X^{ij}$ which agrees with
the version used by Deser [5] in Ricci-flat spaces and written as $L( T_{ab})\equiv
\nabla_i \nabla^i T_{ab} -2 R_{aibj}T^{ij}$.
 (In [5] and [6] a
different sign convention is used for the Riemann tensor than we use in this paper, following
[1], so in the counterparts to these expressions  in [5], [6], the terms involving
Riemann tensors have a sign difference.)}; from
this it follows that
$\square_LT_{ab}$ is divergence-free, $\nabla^b \bigl(\square_L T_{ab}\bigr)=0$.
 This shows that  part (a) of the BES result
can be seen as    a consequence of the fact that
$T_{ab}$ is divergence-free, combined with  
the commutativity property (4).

An obvious question is whether such a rearrangement as (3) and a commutator   property such
as (4) also exist in spaces more general than Ricci-flat spaces.  In this letter we
show that, with an alternative generalised wave operator
$\square_G$,  we can obtain an analogous --- and we argue deeper --- result for
{\it Einstein-Maxwell spaces}.

\medskip

We  consider directly   $H_{ab}$ from (2), and rearrange as follows:
$$ \eqalign{2 H_{ab} & =     \Bigr(\square\bigl(F_{ac}F_b{}^c\bigr) -
F_{ac}\square 
F_b{}^c -
F_b{}^c\square 
F_{ac} - {1\over 4} g_{ab}\square\bigl(F_{cd}F^{cd}\bigr) + {1\over 2} g_{ab}F_{cd}\square 
F^{cd}\Bigl)   
\cr & =  
\square\Bigl(F_{ac}F_b{}^c - {1\over 4} g_{ab}F_{cd}F^{cd}\Bigr)  
\cr &
\quad -F_a{}^c \Bigl(2C_{ijbc}F^{ij}+ 2 {n-4 \over n-2}F_{i[b}\tilde R_{c]}{}^i - {2(n-2)\over
n(n-1)}F_{bc}R\Bigr)
\cr &
\qquad -F_b{}^c \Bigl(2C_{ijac}F^{ij}+ 2 {n-4 \over n-2}F_{i[a}\tilde R_{c]}{}^i - {2(n-2)\over
n(n-1)}F_{ac}R\Bigr)
\cr &
\qquad\quad + {1\over 2} g_{ab}F^{cd} \Bigr(2C_{icjd}F^{ij}+ 2 {n-4 \over n-2}F_{i[c}\tilde
R_{d]}{}^i - {2(n-2)\over n(n-1)}F_{cd}R\Bigl) \
}
\eqno(5)$$
having made use of the result which follows from the  the source-free Maxwell's equations, that
[7]
$$\square F_{ab} =
2C_{ijab}F^{ij}+ 2 {n-4 \over n-2}F_{i[a}\tilde R_{b]}{}^i - {2(n-2)\over
n(n-1)}F_{ab}R  
\eqno(6)$$
where $\tilde R_{ab} (\equiv R_{ab} -R g_{ab}/n)$ is the trace-free part of the Ricci tensor
$R_{ab}$. 

When we specialise to {\it four} dimensions, not only do the
terms with the trace-free Ricci tensor disappear identically in (5), but we can exploit the
{\it
four-dimensional identity}
$$C_{[ab}{}^{[de}\delta_{c]}^{f]} = 0\eqno(7)$$ 
to obtain the identity
$$0=9F^{ij} F_{kl} C_{[ai}{}^{[bk}\delta_{j]}^{l]}
=-2 C_{ij}{}^{b}{}_{c}F^{ij}F_a{}^c-2C_{ijac}F^{ij}F^{bc}  +
\delta_a^b F^{cd} C_{ijcd}F^{ij} + 4 C_{ai}{}^{bk} F^{ij}F_{kj} 
\eqno(8)$$
and hence (5) becomes, in  {\it four-dimensional spaces},
$$2 H_{ab} = \square T_{ab} - 2
T_{ij} C_{a}{}^i{}_b{}^j    + {2R\over 3} T_{ab}.
\eqno(9)$$
Note that in calculating (9) we have used the source-free Maxwell equations and the definition
of the electromagnetic energy-momentum tensor in (1), but not Einstein's equations. 

Although this expression for $H_{ab}$ looks similar to the expression (3) exploited by Deser
[5]  and written in terms of the  Lichnerowicz  operator
$\square_L $ for  Ricci-flat spaces, this version (9)  does
not coincide with
$\square_L T_{ab}/2$ for {\it arbitrary} spaces.

For {\it Ricci-flat   spaces}, the divergence of  $\square_L X_{ab}$  (or, in
particular, of $H_{ab}=\square_L
T_{ab}/2$) follows directly from the divergence of arbitrary $X_{ab}$ (or, in particular, of
$T_{ab}$). On the other hand, for the full non-linear case, when we calculate the divergence
of $H_{ab}$ in (9), we find that we need to use explicitly the fact that {\it $T_{ab}$ is a
divergence-free tensor which is equivalent to the Einstein tensor via Einstein's equations.} 
We could show this by a direct calculation of the divergence of
$H_{ab}$ from (9), but it will be more instructive to exploit a commutator property,
motivated by the special commutator identity in [5]   for $n$-dimensional 
Ricci-flat spaces. 
Any symmetric $2$-tensor $X_{ab}$ is easily seen to satisfy 
$$\eqalign{\square\nabla^bX_{ab}  
= \nabla^b\Bigl(\square 
X_{ab} +R_{b}{}^{i}X_{ai} - 2 R_{a}{}^i{}_b{}^jX_{ij} 
\Bigr) - 2 \nabla_{[j}R_{a]i}X^{ij}
\ , }\eqno(10a)$$
or equivalently
$$ \Bigl(\delta^i_a\square -  R_{a}{}^{i}\Bigr) \nabla^b X{}_{ib}
=\nabla^b\Bigl(\square 
X_{ab}  -2R^{i}{}_{[a}X_{b]i} - 2 R_{a}{}^i{}_b{}^jX_{ij}  
\Bigr) + \nabla_a R^{ij}X_{ij} \ . \eqno(10b)$$ 

Note that in arbitrary spaces we cannot deduce the existence of divergence-free tensors
involving second derivatives of an arbitrary tensor $X_{ab}$ from either (10a) or (10b);
specialising $X_{ab}$ to be divergence-free but otherwise arbitrary does not help. On the
other hand, there are cases where the existence of  such tensors can be deduced by either
restricting the background spaces (e.g. Ricci-flat spaces as in [5]), and/or restricting the
arbitrariness of the tensor
$X_{ab}$.

If we now replace the arbitrary tensor $X_{ab}$ with the Einstein tensor $G_{ab}
(\equiv R_{ab}-Rg_{ab}/2)$ we obtain, for {\it all $n$-dimensional  spaces},
$$\eqalign{\square_G \nabla^b G_{ab} & 
 = \nabla^b\Bigl(\square 
G_{ab}  -2R_{i[a}G^i{}_{b]} - 2 R_{aibj}G^{ij}  
\Bigr) + \nabla_{a} R_{ij}G^{ij} 
\cr & = \nabla^b\Bigl(\square 
G_{ab}   - 2 R_{aibj}G^{ij}  + 
  {1\over 2} g_{ab}G_{ij}R^{ij} \Bigr)}\eqno(11)$$ 
which yields the commutator identity 
$$\Bigl[\nabla^b,\ \square_G\Bigr]G_{ab}=0\eqno(12)$$
where $\square_G
G_{ab} \equiv 
\square 
G_{ab}  - 2 R_{a}{}^i{}_b{}^jG_{ij}+g_{ab}G_{ij}R^{ij}/2$ and $\square_G
\bigl(\nabla^bT_{ab}\bigr) \equiv \square
\bigl(\nabla^bT_{ab}\bigr)-R^{ai}\nabla^bT_{ab}$; as a consequence of the
divergence-free property of
$G_{ab}$  the  symmetric tensor
$\square_G G_{ab}$ is divergence-free.

We note that this tensor $\square_G G_{ab}$ is a purely geometric construction which is 
divergence-free in {\it all} $n$-dimensional spaces; it has previously been identified in a
number of different papers [8].

Using Einsteins's equations $(G_{ab}=-\kappa T_{ab})$ to replace
the Einstein tensor with the energy-momentum tensor, and  specialising   to four
dimensions, we  obtain the divergence-free tensor
$$\eqalign{\square_G T_{ab}& = \square  
T_{ab} - 2 C_{aibj}T^{ij}  -2\kappa\Bigl(T_{ia}T^i{}_b
-{1\over 4} T^{ij}T_{ij}g_{ab}\Bigr)    +\kappa{2T\over 3}\Bigl(
T_{ab} -Tg_{ab}/4\Bigr)} \ .
\eqno(13)$$
A further specialisation to electromagnetic fields means that the penultimate term in (13) 
disappears when we use the four-dimensional algebraic Rainich
conditions [9], and the final term disappears when we substitute  $T=0$,  resulting in  the
divergence-free tensor,
$$\eqalign{\square_G{T}_{ab}& = \square  
T_{ab} - 2 C_{aibj}T^{ij} }   
\eqno(14)$$

When we apply Einstein's equations  in (9)
(substituting  R=0), and compare with equation (14), we obtain confirmation  of part (b) of
the BES result, and in particular that $ H_{ab}=\square_G T_{ab}/2$ is divergence-free in
Einstein-Maxwell spaces in four dimensions.  (In [2] the result for Einstein-Maxwell spaces
was shown to be valid even for the case where Einstein's equations had a non-zero cosmological
constant; this can easily be confirmed by comparing the coefficients of the $T$ terms in (9)
and (14) respectively.)

\medskip

The original BES results were obtained in spinors, and the authors pointed out that a
tensor derivation was 'far from obvious'. The inbuilt four-dimensional simplicity for spinors
has its counterpart in the much more complicated exploitation of four-dimensional identities in
tensors [10].   In the
above analysis it is clear that the
four-dimensional identity (7) plays a crucial role in the tensor proof of part (b) of the BES
result (as it also does in part (a)); in addition the
four-dimensional Rainich identity is essential to simplify (5) in the proof of part (b). 
For {\it higher dimensions $n>4$}, it is easy to see that the terms involving the trace-free
Ricci tensor $\tilde R_{ab}$ do not disappear in (5), and  it is straightforward to show that
there are no analogous identities  which would enable the quadratic terms in
$F_{ab}$  on the right hand side of (5) to  be replaced with
$T_{ab}$, nor are there any  appropriate higher dimensional   Rainich identities.
Hence the divergence-free  property of $H_{ab}$ is restricted to {\it four}
dimensions in Einstein-Maxwell spaces, as is the identity (9).

\medskip

These calculations demonstrate  formally, as in flat and Ricci-flat spaces, that the
divergence-free nature of the energy-momentum tensor $T_{ab}$ leads to the divergence-free
property of an expression which can be considered as a generalised wave operator on
$T_{ab}$. In the flat space case Deser [5] argued that this expression  supplied no independent
content, and the same criticism might be made of the Ricci-flat space case --- in both of
these cases the divergence of {\it any} tensor $X_{ab}$ leads {\it directly} to the
divergence of a generalised wave operator on
$X_{ab}$. However, in Einstein-Maxwell spaces there is a significant new input required to
ensure the divergence of $H_{ab}=\square_GT_{ab}$ --- Einstein's equations are crucial. A
confirmation of the fact that  the divergence-free property  of $H_{ab}$ brings
something new is given by Senovilla's demonstration that non-trivial divergence-free currents
can be constructed from $H_{ab}$ in situations where the divergence-free
currents constructed from $T_{ab}$ are trivial [11].

Although the general symmetric divergence-free tensor $\square_G G_{ab}$ constructed from
the divergence-free Einstein tensor $G_{ab}$ has been quoted in a number of places in the
literature [8] (usually as an aside in the investigations of $4$-index tensors of the
Bel-Robinson type [12]), there seems to have been no significant investigation of the
energy-momentum counterpart  $\square_G T_{ab}$ obtained via Einstein's equations. The
significance and  usefulness of this tensor and the corresponding divergence-free currents
need to be explored.

\

\

{\bf Acknowledgements.}

Thanks to Jos\'{e} Senovilla, G\"oran Bergqvist, Ingemar Eriksson, 
Anders H\"oglund  and Magnus Herberthson for discussions and comments, and to Jos\'{e} Senovilla
for pointing out the links with references [8]. I also wish to acknowledge the ongoing
financial support of  Vetenskapsr\aa det (the Swedish Research Council).

\

\

{\bf References.}

[1] R. Penrose and W. Rindler, {\it Spinors and space}, 2 vols.
(Cambridge Univ. Press, Cambridge, 1984, 1986).

[2] G. Bergqvist, I. Eriksson, J. M.
M. Senovilla,  {\it Class. Quan. Grav.}, {\bf 20}, 2663
(2003).

[3]  M. Chevreton, {\it Nuovo Cimento}, {\bf 34}, {901}, (1964).

[4]  J. M. M. Senovilla, 
      {\it Class. Quan. Grav.},
     {\bf 17}, 2799 (2000); J. M. M. Senovilla,  {\it Gravitation and Relativity in
General}, eds. A. Molina, J. Mart\'{\i}n, E.Ruiz and F. Atrio
(World Scientific, 1999), (Preprint gr-qc/9901019).

[5] S. Deser, {\it Class. Quan. Grav.}, {\bf 20}, L213,
(2003).

[6] A. Lichnerowicz, {\it Publ. Math. IHES,} {\bf 10}, 293. (1961)

[7] F. Andersson and S. B. Edgar, {\it Int. J. of Mod. Phys. D}, {\bf 5}, 217, (1996).

[8] 
 C. Gregory, {\it Phys. Rev.}, {\bf 72}, 72 (1947). \ C. D.  Collinson, {Proc. Camb. Phil.
Soc.}, {\bf 58}, 346. (1962). \
 I. Robinson, 
 {\it Class. Quant Grav.}, {\bf 14}, {A331}, (1997). \
 A. Balfag\'on and X. Ja\'en,  {\it Class. Quant Grav.}, {\bf 17},
{2491}, (2000).

[9] G. Y. Rainich, {\it Trans. Am. Math. Soc.}, {\bf 27}, 106, (1925).

[10] S. B. Edgar and  Ola Wingbrant, "Old and new results for superenergy
tensors from  dimensionally dependent tensor identities."  Preprint
gr-qc/0304099. To be published in J. Math. Phys. December, 2003.

[11]  J. M. M. Senovilla, "New conservation laws for electromagnetic fields in gravity".
Private communication.

[12]  L. Bel, {\it CR Acad. Sci. Paris}, {\bf 246}, {3105}, (1958);
{\it CR Acad. Sci. Paris}, {\bf 247}, {1094}, (1958); {\it CR Acad. Sci. Paris}, {\bf 248}, {1297}, (1959); 
 I. Robinson, unpublished King's College Lectures (1958);
 {\it Class. Quant Grav.}, {\bf 14}, {A331}, (1997)
\end